\documentstyle[12pt]{article}
\def\be{\nopagebreak[3]\begin{equation}}
\newcommand{\ee}{\end{equation}}
\newcommand{\rf}[1]{Ref.~\cite{#1}}
\newcommand{\eq}[1]{Eq.~(\ref{#1})}
\renewcommand{\pd}[2]{\frac{\partial #1}{\partial #2}}
\newcommand{\arcos}{{\rm arcos}\,}
\newcommand{\im}{{\rm Im}\,}
\newcommand{\re}{{\rm Re}\,}
\newcommand{\tr}{{\rm tr}\,}
\newcommand{\lam}{\lambda}

\newcommand{\pip}{\pi_{\scs{+}}}
\newcommand{\pim}{\pi_{\scs{-}}}

\def\ba{\begin{array}}
\def\ea{\end{array}}

\def\ni{\noindent}
\newcommand{\scs}[1]{{\scriptscriptstyle #1}}

\newcommand{\F}{{\scriptscriptstyle F}}

\begin{document}
\begin{titlepage}
\begin{flushright}
NBI-HE-93-55\\
September 1992\\
\end{flushright}
\vspace*{36pt}
\begin{center}
{\huge \bf
Critical scaling in the matrix model on a Bethe tree}
\end{center}
\vspace{2pc}
\begin{center}
 {\Large D.V. Boulatov}\\
\vspace{1pc}
{\em The Niels Bohr Institute\\
University of Copenhagen\\
Blegdamsvej 17, 2100 Copenhagen \O \\
Denmark}
\vspace{2pc}
\end{center}
\begin{center}
{\large\bf Abstract}
\end{center}
\bigskip

The matrix model with a Bethe-tree embedding space (coinciding at large
$N$ with the Kazakov-Migdal ``induced QCD'' model \cite{KM})
is investigated. We further elaborate the Riemann-Hilbert approach
of \rf{Mig1}  assuming
certain holomorphic properties of the solution. The critical scaling (an
edge singularity of the density) is found  to be $\gamma_{str} =
-\frac{1}{\pi} \arcos D$, for $|D|<1$, and $\gamma_{str} =
-\frac{1}{\pi} \arcos \frac{D}{2D-1}$, for $D>1$.
Explicit solutions are constructed at $D=\frac{1}{2}$ and $D=\infty$.
\vfill
\end{titlepage}

\newpage

\section{Introduction}

Let us start with the matrix model describing surfaces embedded into an
arbitrary graph $\cal G$ defined by its incidence matrix $G_{xy}$
($G_{xy}=1$, if there is a link between
vertices $x$ and $y$, and 0, otherwise). The standard action is

\be
S=-N \sum_{xy}G_{xy}\tr\Big[\Phi(x)\Phi(y)\Big]+
N\sum_x \tr V(\Phi(x))
\label{action}
\ee
where $\Phi(x)$ is an $N\times N$ hermitian matrix attached to an $x$'th
vertex; the potential, $V(\varphi)$, is an arbitrary polynomial.
The partition function is defined as the integral over all field configurations

\be
Z=\int \prod_{x\in \cal G} d\Phi(x)\; e^{-S}
\label{Z}
\ee

At each vertex, $\Phi(x)$ can be decomposed into diagonal, $\varphi(x)$,
and angular, $S(x)\in U(N)$, parts:

\be
\Phi(x)=S^+(x)\varphi(x)S(x)
\label{decomp}
\ee
It is convenient to introduce the on-link gauge variables

\be
\Omega_{xy}=S^+(x)S(y)
\label{variable}
\ee
obeying the constraint that their product along any loop equals unity.

If the graph $\cal G$ is a tree, all the gauge variables are independent
and can be integrated out by the Itzykson-Zuber
formula \cite{ItZub}:

\be
{\cal I}(\phi,\psi)=
\int  d\Omega \: e^{Ntr\phi \Omega\psi \Omega^+} = N^{-N(N-1)/2}
\prod_{n=1}^{N-1}n!\frac{\det_{ab}
e^{ N\phi_a\psi_b}}{\Delta(\phi)\Delta(\psi)}
\label{ItZub}
\ee

\ni
In eq. (\ref{ItZub}), $\phi$ and $\psi$ without loss of generality are
real and diagonal; \\ $\Delta(\phi)=\prod_{i<j}(\phi_i-\phi_j)$
is the Van-der-Monde determinant.

As a result, one finds a model in which a role of dynamical variables is
played by $N$ eigenvalues of $\Phi(x)$.
In the $N\to \infty$ limit, the saddle-point approximation is exact and,
in this sense, the model is soluble.

If the graph $\cal G$ has loops, the model becomes very complicated, and
so far no reliable results have been obtained in this case. Technically,
mean field does not work; physically, models of this kind describe the
interaction of $2d$ gravity with matter having the central charge $c>1$.

In the simplest case of the $c=1$ model compactified on a ring, the
angular degrees of freedom describe vortices, {\em i.e.},
configurations with holes in the string world sheet wrapped along the
ring. By dropping the constraint that the product of the angular
variables along the ring equals unity, one discards the vortices
\cite{GK}. Hence, the singlet (vortex free) sector of the model is
physically sensible. By dropping the constraints in the matrix model
embedded in the regular $D$-dimensional lattice, Kazakov and Migdal
\cite{KM}
obtained the model (sometimes called ``induced QCD'' \cite{Mig2})
generalizing the
$c=1$ vortex-free matrix model. It can be easily seen that, in the
$N\to\infty$ limit, it is equivalent to the matrix model with a
Bethe-tree (BT) target space \cite{Boul}. A Bethe tree is, by definition, an
infinite tree having the same even coordination number, $2D$, at each
vertex. It is the infinite simply connected covering of the regular
$D$-dimensional lattice. Therefore, the BT matrix model can be regarded
as the mean field approximation for lattice scalar field theory.

\section{The matrix model on a Bethe lattice}

In this paper we consider the matrix model with a BT target
space. In the large $N$ limit,
the ground state should be homogeneous. The free
energy per lattice site is given by the matrix integral

\be
F=\log \int d^{\scs{N^2}}X\; e^{-N\tr V(X)}
\Big[I(X)\Big]^{2D}
\label{parfun}
\ee
where $I(X)$ obeys the equation

\be
I(X)=\int d^{\scs{N^2}}Y\; e^{N\tr(XY- V(Y))}
\Big[I(Y)\Big]^{2D-1}
\label{homcon}
\ee
In terms of eigenvalues of $X$ we find

\be
F=\log \int
\prod_{n=1}^{N}dx_n\Delta^2(x)\;
e^{-N\sum_i^N V(x_i)}
\Big[I(X)\Big]^{2D}
\label{parfun2}
\ee

The saddle-point equation for this integral reads

\be
\frac{2}{N}\sum_{j\neq i}\frac{1}{x_i-x_j} - V'(x_i)
+2D\, w_i = 0
\label{spe0}
\ee
where
\be
w_k=\frac{1}{N}\pd{\ }{x_k}\log I(X).
\label{w}
\ee

Let us assume that, in the $N\to \infty$ limit, the eigenvalues
distribution

\be
\rho(x)=\lim_{N\to\infty}\frac{1}{N}\sum_{n=1}^{N}\delta(x-x_n)
\label{distrib}
\ee
has a finite connected support. Then we can introduce the function

\be
f(x)=\int dy \frac{\rho(y)}{x-y}
\label{f(x)}
\ee
where the integral goes over the support of $\rho(y)$ ($Supp\ \rho(y)$).
$f(x)$ is holomorphic
at the infinity $f(x)=\frac{1}{x}+0\big(\frac{1}{x^2}\big)$ and
has one cut on the real axis where

\be
\im f(x)=\pi\rho(x)
\ee

In terms of $f(x)$, \eq{spe0} takes the form

\be
2\re f(x) - V'(x) + 2Dw(x)=0
\label{spe}
\ee
which is valid, strictly speaking, only on $Supp\ \rho(x)$.

\be
w(x)=\pd{\ }{x}\frac{\delta\ \ }{\delta\rho(x)}
\lim_{N\to\infty}\frac{1}{N^2} \log I(X)
\ee
is the large $N$ limit of the function (\ref{w}).

To close the system of equations, we should calculate $w(x)$ as a
functional of $f(x)$, {\em i.e.}, to solve \eq{homcon}. However, we can
avoid doing these complicated calculations. As was shown in \rf{Mig1}
(see also \cite{Boul}),
in the $N\to\infty$ limit, the following equation holds

\be
1=(z-w(x))W(z,x)-\int dy\rho(y)\frac{W(z,x)-W(z,y)}{x-y}
\label{eqw}
\ee
where $W(z,x)$ is the $N\to\infty$ limit of the mean value of
diagonal elements of the resolvent matrix

\be
W_k(z) = \frac{1}{I(X)}\bigg[\frac{1}{z-\frac{1}{N}\pd{\ \ }
{X^{\hbox{}^+}}}\bigg]_{kk}I(X)
\ee
The large $z$ expansion of this function has the form
\be
W(z,x)=\frac{1}{z}+\frac{w(x)}{z^2}+\ldots
\ee

As long as the ground state is considered, the following
homogeneity condition holds

\be
\int dx\;\rho(x)W(z,x) = f(z)
\ee

\eq{eqw} is valid only on $Supp\ \rho(x)$. However, all functions
can be analytically continued into the whole complex plane.
Following \rf{Mig1}, let us introduce the function

\[
F(z,x)=1-\int dy\frac{\rho(y)W(z,y)}{x-y}\equiv 1 -
\lim_{N\to\infty}e^{-\F}\int d^{\scs{N^2}}X\: d^{\scs{N^2}}Y\;
\]\be
\frac{1}{N}\tr \bigg[\frac{1}{z-X}\frac{1}{x-Y}\bigg]
e^{N\tr(XY-V(X)-V(Y))}\Big[I(X)I(Y)\Big]^{2D-1}
\ee
where the integral goes over  $Supp\ \rho(x)$.
By construction, $F(z,x)$ is symmetric $F(z,x)=F(x,z)$.

At $z$ sufficiently large, $F(z,x)$ as a function of $x$ is
holomorphic everywhere except the support, where its imaginary part equals

\be
\im F(z,x)=-\pi\rho(x)W(z,x)
\ee
The real part can be determined from \eq{eqw}

\be
\re F(z,x)=(z-w(x)-\re f(x))W(z,x)
\ee

As $F(z,x)$  is real on the real axis out of the cut, we find
from the Riemann-Schwartz principle that
\be
F(z,\overline{x})=\overline{F}(z,x)
\ee
and, hence, the ratio of limiting values of $F(z,x)$ above and below the
cuts is independent of $W(z,x)$

\be
\frac{F(z,x+i0)}{F(z,x-i0)}=\frac{z-w(x)-\re f(x) - i\im f(x)}
{z-w(x)-\re f(x) + i\im f(x)}
\ee
At the infinity we have by construction

\be
F(z,x)=1-\frac{f(z)}{x}+0\Big(\frac{1}{x^2}\Big)
\label{expF}
\ee

The function having all needed properties is given by the formula

\be
F(z,x)=\exp \oint_C\frac{dy}{2\pi i}\frac{1}{y-x}
\log [z-w(y)-f(y)]
\label{F=}
\ee
where the contour $C$ encircles the cut of $f(y)$ in the positive direction
leaving aside all other singularities (whatever they are), that is,
$F(z,x)$ is defined perturbatively by its expansion in inverse powers of
$z$ and $x$.

For \eq{F=} to make sense, the function $u(x)=w(x)+f(x)$ has to be
holomorphic. However, $w(x)$ depends through the saddle-point equation
(\ref{spe}) on $\re f(x)$. Hence, we should be able to introduce two
holomorphic functions $f_1(x)$ and $f_2(x)$ such that, on $Supp\
\rho(x)$, $f_1(x) = \re f(x)$ and $f_2(x) = \im f(x)$. Off the support,
both $f_1(x)$ and $f_2(x)$ are holomorphic ({\em i.e.}, obey
the Cauchy-Riemann equations separately); on it $\im f_1(x)=\im
f_2(x)=0$.  Everywhere in the complex plane $f(x)=f_1(x)+if_2(x)$, and
we define

\be
u(y)=\frac{1}{2D}V'(y)+\frac{D-1}{D} f_1(y)+i f_2(y)
\ee

Let $Supp\ \rho(x)=(a,b)$, then we can rewrite \eq{F=} as

\[
F(z,x)=\exp -\int_a^b\frac{dy}{2\pi i}\frac{1}{y-x}\log
\bigg[\frac{z-u(y)}{z-\bar{u}(y)}\bigg]=
\]\be
=\exp \int_a^b\frac{du}{2\pi i}\frac{1}{u-z}\log
\bigg[\frac{x-y(u)}{z-\bar{y}(u)}\bigg]
\label{F=2}
\ee
where $y(x)$ is the inverse function: $u(y(x))=x$.
For $F(z,x)$ to be symmetric, $F(z,x)=F(x,z)$, the following remarkable
equation has to hold

\be
y(x)=\bar{u}(x)
\label{y=u}
\ee
or, equivalently,

\be
u(\bar{u}(x))=x
\label{uu=x}
\ee
This equation has previously been obtained by Luochen (unpublished)
and rederived by
Matytsin within a quite different approach \cite{Mat}.

In general, the inverse function $y(x)$ is not unique. However, we are
interesting in critical behavior consisting in the divergence of a
perturbative expansion in a non-gaussian part of the potential $V(x)$.
In the gaussian case, $V_0(x)=\frac{m^2}{2}x^2$, the solution was found by
D.Gross \cite{Gross}:

\be
f(x)=\frac{a}{2}x+i\sqrt{a-\frac{a^2x^2}{4}}
\label{semicirc}
\ee
with 2 possible values of $a$

\be
a_{\scs{\pm}}=\frac{1}{2D-1}\Big[m^2(D-1)\pm
D\sqrt{m^4-4(2D-1)}\Big]
\label{a=}
\ee
The condition that a solution has to describe perturbations
around this semi-circular saddle-point fixes a branch of $y(x)$
unambiguously.
\bigskip

{}From the expansion (\ref{expF}), we find Master Field Equation in the
form

\be
f(z)=\oint_C\frac{dy}{2\pi i}\log [z-w(y)-f(y)]
\label{MFE}
\ee
which is satisfied automatically, if \eq{uu=x} holds.

We can as well obtain the equation

\be
f(z)=-\oint_C\frac{dy}{2\pi i}
\log \Big[z-w(y)- \bar{f}(y)\Big]
\label{MFEdual}
\ee

then, obviously,

\be
\oint_C\frac{dy}{2\pi i}
\log \Big[\Big(x-\frac{1}{2D}V'(y)-
\frac{D-1}{D} f_1(y)\Big)^2
+\Big(f_2(y)\Big)^2\Big]=0
\label{int=0}
\ee
and we conclude that

\be
f_2(x)=\sqrt{\Phi(x)}
\ee
where $\Phi(x)$ is holomorphic inside the contour $C$.
Critical behavior takes
place when (i) a branching point or (ii)
a zero of $\Phi(x)$ approach the support
of the eigenvalues density (its edges correspond to zeros of
$\Phi(x)$), in other words, when two singularities pinch the contour.
In the first (i) case, the order of the branching point
increases, which is characteristic of $c<1$ models. The second
(ii) possibility
corresponds to the shrinking of a handle of the Riemann surface of
$f(x)$, as in the $c=1$ matrix model.

\section{Critical scaling}

In matrix models, critical behavior is determined by a scaling near an
edge of the eigenvalues density exactly at a critical point, when

\be
\rho(x)\propto (-x)^{1+\gamma}\hspace{1cm} x<0
\ee
where $\gamma\equiv -\gamma_{str}>0$ is a critical exponent (for
simplicity, we have assumed that the branching point of $f(x)$ is at
$x=0$). In the small vicinity, $|x|\ll 1$, we can expand all functions in
powers of $x$. For example, (if $0<\gamma<1$)

\be
u(x) \approx ax +
b\Big(\frac{D-1}{D}\cos\pi\gamma+i\sin\pi\gamma\Big)
e^{-i\pi\gamma}x^{1+\gamma}+\ldots
\ee
and
\be
\bar{u}(x) \approx ax +
b\Big(\frac{D-1}{D}\cos\pi\gamma-i\sin\pi\gamma\Big)
e^{-i\pi\gamma}x^{1+\gamma}+\ldots
\ee
Expanding \eq{uu=x}, we find

\be\ba{l}
x=\bar{u}(u(x)) \approx a^2x+abe^{-i\pi\gamma}x^{1+\gamma}
\Big[\frac{D-1}{D}\cos\pi\gamma+i\sin\pi\gamma
+\Big(\frac{D-1}{D}\cos\pi\gamma-i\sin\pi\gamma\Big)a^\gamma\Big]
\ea\ee
Hence, $a^2=1$. If we take $a=1$, $a^\gamma=1$, the only solution
is $\gamma=\frac{1}{2}$, which is the well-known one-matrix-model
exponent. Indeed, this solution should always be possible. For example,
at $D=\frac{1}{2}$, it corresponds to the disordered phase of the Ising
model.

The second possibility is $a=-1$, which can be interpreted as a
quadratic top of the effective potential,
$V_{eff}=\frac{1}{2D}V'(x)+\frac{D-1}{D}f_1^{reg}(x)$, where
$f_1^{reg}(x)$ is a regular at $x=0$ part of $f_1(x)$. Therefore, one
should expect critical behavior in this case. In principle, we can
take 2 branches of the function $a^{\gamma}$. In the first case,
$a^\gamma=e^{i\pi\gamma}$, we find the equation

\be
\cos\pi\gamma = D
\label{gamma1}
\ee
which makes sense for $|D|<1$ and fits 3 exactly soluble cases:
$D=0,\frac{1}{2}$ and 1.

The second possibility is $a^\gamma=e^{-i\pi\gamma}$. In this case, a
simple algebra gives the equation

\be
\cos\pi\gamma = \frac{D}{2D-1}
\label{gamma2}
\ee
which formally makes sense for $D>1$ and $D<\frac{1}{3}$.

\section{Explicit solution at $D=\frac{1}{2}$ and $\infty$}

In the previous section we have determined possible types of
singularities of $\rho(x)$. However, it would be instructive to have a
non-trivial global solution to the problem.

Let us suppose that there exist holomorphic coordinates $\lam$ in which the
function $f(x)$ becomes an integral function:

\be
f(x(\lam))\equiv\varphi(\lam)=\sum_{k\geq 1}\varphi_k \lam^k
\ee
and the infinity of the $x$-plane is mapped to the center of the
$\lam$-plane ({\em i.e.,} $x\sim\frac{1}{\lam}$, when $\lam\to 0$). It
means that, geometrically, the Riemann surface of $f(x)$ is a sphere with
one puncture.

There should exist an involution
transformation of the $\lam$-plane, $\pi(\lam)$, preserving a curve of
real $x$'s,

\be
\pi(\pi(\lam))=\lam \hspace{2cm} x(\pi(\lam))=x(\lam),
\ee
such that, on the curve (and only on it),
the action of $\pi$ coincides with the complex conjugation:

\be
\overline{\varphi(\lam)}=\varphi(\pi(\lam))
\ee
Then, as $f(\bar{x})=\bar{f}(x)$, we can identify
\be\ba{l}
\varphi_1(\lam)\equiv f_1(x(\lam))=
\frac{1}{2}[\varphi(\lam)+\varphi(\pi(\lam))] \vspace{1pc} \\
\varphi_2(\lam)\equiv f_2(x(\lam))=
\frac{1}{2i}[\varphi(\lam)-\varphi(\pi(\lam))]
\ea\ee
The functions $u(x)$ and $\bar{u}(x)$ define an invertible map of the
$\lam$-plane onto the $\mu$-plane and back, correspondingly, determined
by the equations

\be\ba{l}
x(\mu)=\frac{1}{2D}V'(x(\lam))+\frac{2D-1}{2D}\varphi(\lam)
-\frac{1}{2D}\varphi(\pi(\lam)) \vspace{1pc} \\
x(\lam)=\frac{1}{2D}V'(x(\mu))-\frac{1}{2D}\varphi(\mu)
+\frac{2D-1}{2D}\varphi(\pi(\mu))
\label{sys}
\ea\ee

The simplest realization of this scheme,

\be
x(\lam)=\frac{1}{\lam}+a\lam\hspace{2cm}\varphi(\lam)=\lam,
\ee
corresponds to the semi-circle solution (\ref{semicirc}). Here,

\be
\pi(\lam)=\frac{1}{a\lam}
\ee

For the gaussian potential, $V'(x)={m^2}x$, equations (\ref{sys}) take
the form

\be\ba{l}
\frac{1}{\mu}+a\mu=\frac{1}{2Da}(m^2a-1)\frac{1}{\lam}
+\frac{1}{2D}(m^2a+2D-1)\lam \vspace{1pc}\\
\frac{1}{\lam}+a\lam=\frac{1}{2Da}(m^2a+2D-1)\frac{1}{\mu}
+\frac{1}{2D}(m^2a-1)\mu
\ea\ee
from which we find

\be
\mu=\frac{2Da}{m^2a-1}\lam
\ee
together with the self-consistency equation for $a$:

\be
(m^2a+2D-1)(m^2a-1)=(2Da)^2
\ee
which gives the solution (\ref{a=}).

Now, let us consider the non-gaussian potential
$\frac{1}{2D}V'(x)=m^2x+gx^2$ and
try the ansatz

\be
x(\lam)=\frac{1}{\lam}+c+a\lam+b\lam^2
\label{ansatz}
\ee
{}From the power counting, we find the form of $\varphi(\lam)$:

\be
\varphi(\lam)=\lam+\varphi_2\lam^2+\varphi_3\lam^3+\varphi_4\lam^4
\ee

In this case, $\pi(\lam)$ obeys the quadratic equation

\be
b\lam\pi^2(\lam)+(a\lam+b\lam^2)\pi(\lam)-1=0
\ee
which has two solutions

\be
\pi_{\scs{\pm}}(\lam)=\frac{1}{2}\Big[-\frac{a}{b}-\lam\pm
\sqrt{\frac{4}{b\lam}+\Big(\frac{a}{b}+\lam\Big)^2}\Big]
\ee
We have to use $\pip(\lam)$ for the conjugation, because it is an
involution: $\pip(\pip(\lam))=\lam$. Its asymptotic behavior is

\be
\pip(\lam)=\left\{\ba{ll}
\frac{1}{\sqrt{b\lam}},&\mbox{when $\lam\to 0$}\\
\frac{1}{b\lam^2},&\mbox{when $\lam\to\infty$}
\ea\right.
\label{asymp}
\ee

It is natural to suppose that zero and the infinity are fixed points of
the function $\mu(\lam)$ ($\mu(0)=0$ and
$\mu(\infty)=\infty$) as it was in the gaussian case. We see then that a
number of terms on the right hand side of \eq{sys} have to cancel
identically, which gives us the asymptotics of $\mu(\lam)$:

\be
\mu(\lam)=\left\{\ba{ll}
\sqrt{g\lam},&\mbox{when $\lam\to 0$}\\
\sqrt{bg}\lam^2,&\mbox{when $\lam\to\infty$}
\ea\right.
\label{asym}
\ee
together with the answer for $\varphi(\lam)$:

\be
\varphi(\lam)=\lam+
[m^2b+g(a^2+2bc)]\lam^2+
2abg\lam^3+
b^2g\lam^4
\ee
If $D\neq\frac{1}{2}$, we find  contradictions and the equations have no
solution. One could have expected it from the very beginning, because
the ansatz (\ref{ansatz}) is well known in the orthogonal polynomials
approach to the two-matrix model and gives
\mbox{$\gamma_{str}=-\frac{1}{3}$} \cite{DKK}.

At this point, \eq{sys} take the form

\be\ba{l}
\frac{1}{\mu}+c+a\mu+b\mu^2 =
b^2g\lam^4+2abg\lam^3+
(m^2b+g(a^2+2bc))\lam^2+ \\
((m^2+2cg)a+bg)\lam+
m^2c+(2a+c^2)g+ \\
\{b^2g\lam^3+abg\lam^2+(m^2+2cg)b\lam +2bg+(m^2+2cg)a-1\}\pip(\lam)
\ea
\label{eq1}
\ee
and

\be
\frac{1}{\lam}+c+a\lam+b\lam^2=
\frac{g}{\mu^2}+(m^2+2gc)\frac{1}{\mu}
+(2a+c^2)g+m^2c+\sqrt{\frac{b}{g}}\mu
\label{eq2}
\ee

Let us assume that the function $\mu(\lam)$ obeys a quadratic equation,
then there is no ambiguity left  and the ansatz having correct
asymptotic behavior is of the form

\be
\mu(\lam)=-\sqrt{bg}\lam\pim(\lam)=
\sqrt{bg}\lam\pip(\lam)+\sqrt{\frac{g}{b}}(a\lam+b\lam^2)
\label{mu(lam)}
\ee
It can be easily checked that this function does satisfy
Eqs.~(\ref{eq1},\ref{eq2}) provided the parameters obey the following
three constraints

\be\ba{l}
a=\sqrt{\frac{b}{g}}(m^2+2gc)\\
gc^2+(m^2-1)c+2ag=0\\
2bg-\sqrt{\frac{b}{g}}+(m^2+2gc)a-1=0
\label{par1}
\ea\ee
which completes the solution in the $D=\frac{1}{2}$ case.

If $D=\infty$, \eq{sys} takes a similar form and the same ansatz
(\ref{ansatz}) should work. In this case, it is
convenient to introduce the function $\lam(\mu)$ which coincides with
\eq{mu(lam)}:

\be
\lam(\mu)=-\sqrt{bg}\mu\pim(\mu)
\label{lam(mu)}
\ee
and, in complete analogy with the previous case, we find

\be
\varphi(\lam)=\lam-
[m^2b+g(a^2+2bc)]\lam^2-
2abg\lam^3-
b^2g\lam^4
\ee
where parameters obey the equations

\be\ba{l}
a=\sqrt{\frac{b}{g}}(m^2+2gc)\\
gc^2+(m^2-1)c+2ag=0\\
2bg-\sqrt{\frac{b}{g}}+(m^2+2gc)a+1=0
\ea
\label{par2}
\ee
The only difference between \eq{par1} and \eq{par2} is the sign before 1
in the last equation.

At the critical point we have

\be
x(\lam)=\frac{(1+s\lam)^3}{\lam}
\ee
and a simple algebra gives

\be
s^2=\left\{\ba{rl}
\frac{1}{10},&\mbox{when $D=\frac{1}{2}$}\\
-\frac{1}{10},&\mbox{when $D=\infty$}
\ea\right.
\ee

The critical coupling, $g_{c}$, is to be determined from the equation

\be
\sqrt{sg_{c}}=1
\ee
Hence, at $D=\infty$, $g_{c}$ is complex.

\section{Discussion}

The main result of this paper, \eq{gamma1}, looks very natural. After
the
identification $D=\frac{n}{2}$, it coincides with the Kostov's solution
of the O($n$) matrix model \cite{Ivan}. It is a standard matrix-model
singularity. The second branch (\ref{gamma2})
corresponds to a perturbation expansion
around a local maximum of the free energy \cite{Gross} (only this
saddle-point of the gaussian model is non-singular at $D=\infty$).
Moreover, we have found that the critical coupling in this case is
complex (at least, at $D=\infty$). This means, presumably, that, at
$D>1$, our model is in the usual sense non-critical as all solved so far
matrix models.

\section{Acknowledgments}

I would like to thank J.Ambj\o rn, V.Kazakov,
Yu.Makeenko and, especially, A.A.Migdal for the discussions.
Financial support from the EEC grant CS1-D430-C is gratefully
acknowledged.

\end{document}